\documentclass[twocolumn,tighten]{aastex61}
\usepackage{amssymb, amsfonts, amsmath,framed}

\usepackage{bm}
\expandafter\ifx\csname package@font\endcsname\relax\else
 \expandafter\expandafter
 \expandafter\usepackage
 \expandafter\expandafter
 \expandafter{\csname package@font\endcsname}
\fi
\expandafter\ifx\csname package@font\endcsname\relax\else
 \expandafter\expandafter
 \expandafter\usepackage
 \expandafter\expandafter
 \expandafter{\csname package@font\endcsname}
\fi
\def\bq{\begin{equation}}
\def\eq{\end{equation}}
\def\bqy{\begin{eqnarray}}
\def\eqy{\end{eqnarray}}

\begin{document}

\title{The dehydration of water worlds via atmospheric losses}

\correspondingauthor{Chuanfei Dong}
\email{dcfy@princeton.edu}

\author{Chuanfei Dong}
\affiliation{Department of Astrophysical Sciences, Princeton University, Princeton, NJ 08544, USA}
\affiliation{Princeton Center for Heliophysics, Princeton Plasma Physics Laboratory, Princeton University, Princeton, NJ 08544, USA}

\author{Zhenguang Huang}
\affiliation{Center for Space Environment Modeling, University of Michigan, Ann Arbor, MI 48109, USA}

\author{Manasvi Lingam}
\affiliation{Harvard-Smithsonian Center for Astrophysics, Cambridge, MA 02138, USA}
\affiliation{John A. Paulson School of Engineering and Applied Sciences, Harvard University, Cambridge, MA 02138, USA}

\author{G\'abor T\'oth}
\affiliation{Center for Space Environment Modeling, University of Michigan, Ann Arbor, MI 48109, USA}

\author{Tamas Gombosi}
\affiliation{Center for Space Environment Modeling, University of Michigan, Ann Arbor, MI 48109, USA}

\author{Amitava Bhattacharjee}
\affiliation{Department of Astrophysical Sciences, Princeton University, Princeton, NJ 08544, USA}
\affiliation{Princeton Center for Heliophysics, Princeton Plasma Physics Laboratory, Princeton University, Princeton, NJ 08544, USA}

\begin{abstract}
We present a three-species multi-fluid MHD model (H$^+$, H$_2$O$^+$ and e$^-$), endowed with the requisite atmospheric chemistry, that is capable of accurately quantifying the magnitude of water ion losses from exoplanets. We apply this model to a water world with Earth-like parameters orbiting a Sun-like star for three cases: (i) current normal solar wind conditions, (ii) ancient normal solar wind conditions, and (iii) one extreme ``Carrington-type'' space weather event. We demonstrate that the ion escape rate for (ii), with a value of 6.0$\times$10$^{26}$ s$^{-1}$, is about an order of magnitude higher than the corresponding value of 6.7$\times$10$^{25}$ s$^{-1}$ for (i). Studies of ion losses induced by space weather events, where the ion escape rates can reach $\sim$ 10$^{28}$ s$^{-1}$, are crucial for understanding how an active, early solar-type star (e.g., with frequent coronal mass ejections) could have accelerated the depletion of the exoplanet's atmosphere. We briefly explore the ramifications arising from the loss of water ions, especially for planets orbiting M-dwarfs where such effects are likely to be significant.
\end{abstract}

\section{Introduction} \label{SecIntro}
The past decades have witnessed the discovery of thousands of exoplanets. Much effort has been devoted to locating rocky exoplanets in the habitable zone (HZ) of their host star, as these planets are theoretically capable of supporting liquid water - an essential ingredient for life-as-we-know-it. However, a remarkable feature of exoplanetary science is that it has opened up the possibility of discovering planets that are very dissimilar to those found in our Solar system. 

A noteworthy example in this category is \emph{ocean planets}, also referred to as water worlds. These planets are anticipated to be volatile-rich \citep{Kuch03}, and possess oceans that are conceivably hundreds of kilometers deep \citep{Leg04}. Notable candidates for water worlds include Gliese 1214 b \citep{Char09}, the six planets in orbit around Kepler-11 \citep{DAB16}, Kepler-62e and Kepler-62f \citep{KSR13}, and Kepler-22b \citep{Bor12}. Recent studies indicate that most super-Earths with radii $\gtrsim 1.6 R_\oplus$ are not typically rocky since they are volatile-rich planets \citep{Rog15}; a stringent upper bound on the radius was also identified by \citet{CK17}. The habitability of ocean planets has been extensively investigated, such as their interior structure and dynamics \citep{Net11}, mass-radius relationships \citep{BS13}, and atmospheric chemistry \citep{KSR13,Gold15}.

We also emphasize that terrestrial planets in our Solar system are believed to have possessed primordial oceans, although the evidence remains disputed. Most notably, there have been several proposals that the Earth was initially an ocean-dominated world \citep{IK10,AN12}, perhaps even up to the late-Archean era at $2.5$ Gya \citep{FCR08}. In addition, many studies have suggested the existence of a northern ocean of liquid water on ancient Mars; the reader may consult \citet{Word16} for more details. Venus could also have had a shallow ocean \citep{Way16} until about $1$ Gya. Thus, on account of all these reasons, water-rich exoplanets merit further study. 

One of the most important factors studied in planetary habitability is the long-term existence of an atmosphere. To this end, it is necessary to quantify the extent of atmospheric losses from planets. When studying the atmospheric escape rates of exoplanets, especially in the context of water loss, it has become common to rely upon models of thermal (Jeans or hydrodynamic) escape. However, as seen from NASA's Mars Atmosphere and Volatile EvolutioN (MAVEN) mission \citep{jakosky15b,brain15}, ion escape rates also play an important role; in fact, for planets larger than Earth, neutral losses are not necessarily the dominant mechanism \citep{DLMC}.

Thus, when modeling atmospheric escape rates, it is necessary to take into account stellar wind-exoplanet interactions, and quantify ion escape rates by using sophisticated magnetohydrodynamic (MHD) models. Such studies have been undertaken recently for Proxima Centauri b \citep{Aira17,DLMC} and the TRAPPIST-1 planets \citep{DJL}. An important to be noted is that these studies implicitly assumed that the atmospheric composition was similar to that of Mars, Venus or Earth. In this paper, we shall consider a scenario wherein the planet's atmosphere is primarily comprised of water vapor \citep{Gold15},\footnote{Although this assumption is not entirely realistic, it could serve as a reasonable approximation for planets like GJ 1214b that are potentially characterized by water-rich atmospheres \citep{MRK12}.} and determine the corresponding ion escape rates.

\section{Multi-fluid MHD Model} \label{SecModDes}
In this Section, we describe the multi-fluid MHD model, endowed with the electron pressure equation, that is used to simulate the ion escape processes on water worlds. We present the model equations and clarify the physics behind certain terms. 

\subsection{The significance of multi-fluid MHD models}
Before embarking on our description of the multi-fluid MHD equations, a brief description of their utility and accuracy is warranted in the context of our Solar system. When we refer to a multi-fluid MHD model, it must be understood that the continuity, momentum and energy equations for \emph{each} species are prescribed, unlike the better-known multi-species single-fluid MHD model where only one equation is prescribed for the momentum and for the energy \citep{Ma04}. Although multi-fluid MHD models are computationally more expensive, they have the advantage of improved realism and accuracy \citep{Toth12,dong14}.

Multi-fluid MHD models have been successfully applied to analyze a wide range of planetary plasma environments in the Solar system. A few notable examples include Mars \citep{najib11,dong14,dong15a}, Earth \citep{Glo09,Bra10}, Europa \citep{rubin15}, Enceladus \citep{Jia12}, Ganymede \citep{PW04} and comets \citep{rubin14b,huang16a,Huang16b}. In all of these instances, the obtained theoretical results from the simulations have been shown to be in very good agreement with observations from missions such as MAVEN, Galileo and Rosetta. For instance, in the case of Mars, it was shown that the multi-fluid MHD equations characterized the ion escape rates from the upper atmosphere more accurately as compared to the multi-species equations \citep{dong14}. 

Thus, in order to model the effects of water ion losses from ocean planets, we shall introduce a novel multi-fluid MHD model which includes two ion fluids and one electron fluid. The model incorporates all of the key chemical reactions between neutral water molecules and charged particles and is therefore well-suited for studying the stellar wind-induced ion escape in ocean planets. The numerical code developed in this paper represents the first self-consistent model in the literature that is capable of estimating the ion escape rates from (exo)planets with water vapor atmospheres.

\subsection{The multi-fluid MHD equations} \label{SSecMFMHD}
We use $\rho$, $\mathbf{u}$, $p$, $\overleftrightarrow{I\,}$, $k_B$ and $ \gamma$ to denote the mass density, velocity vector, pressure, the identity matrix, the Boltzmann constant and the specific heat ratio, respectively. The multi-fluid MHD code simulates three fluids in total. Of this trio, two of them are ion fluids - water group ions $\rm H_2O^+$ and stellar wind protons $\rm H^+$ - that are distinguished via the subscript $s$. The third is the electron fluid that is denoted by the subscript $e$; note that we also introduce the subscript $n$ for neutral background species. The multi-fluid MHD equations are listed as follows \citep{najib11,rubin14b}:

\begin{eqnarray} \label{ionmass}
&& \frac{\partial \rho_{s}}{\partial t}+\nabla\cdot(\rho_{s}\mathbf{u_{s}})=\mathcal{S}_{s}-\mathcal{L}_{s} \\ \nonumber \label{ionmom}
&&\frac{\partial \left(\rho_{s}\mathbf{u_{s}}\right)}{\partial t}+\nabla\cdot\left(\rho_{s}\mathbf{u_{s}u_{s}}+p_{s}\overleftrightarrow{I\,}\right)=n_{s}q_{s}\left(\mathbf{u_{s}-u_{+}}\right)\times\mathbf{B} \\ \nonumber
&& \hspace{0.5 in} +\frac{q_{s}n_{s}}{en_{e}}\left(\mathbf{J}\times \mathbf{B}-\nabla\mathit{p_{e}}\right)+\rho_s\mathbf{G}\\ 
&& \hspace{0.5 in} +\rho_{s}\sum_{t=\mathrm{all}}\nu_{st}(\mathbf{u_{t}-u_{s}})+\mathcal{S}_{s} \mathbf{u_{n}}-\mathcal{L}_{s} \mathbf{u_{s}} \\  \nonumber \label{ionpress}
&&\frac{\partial p_{s}}{\partial t}+\left(\mathbf{u_{s}}\cdot\nabla\right)p_{s}=-\gamma_{s} p_{s}\left(\nabla\cdot\mathbf{u_{s}}\right) \\ \nonumber
&&\hspace{0.4 in} + \sum_{t=\mathrm{all}}\frac{\rho_{s}\nu_{st}}{m_{s}+m_{t}}\left[2k_B\left(T_{t}-T_{s}\right)+\frac{2}{3}m_{t}\left(\mathbf{u_{t}-u_{s}}\right)^{2}\right] \\ 
&& \hspace{0.4 in} + k_B\frac{\mathcal{S}_{s}T_{n}-\mathcal{L}_{s}T_{s}}{m_{s}}+\frac{1}{3}S_{s}\left(\mathbf{u_{n}-u_{s}}\right)^{2} \\ \nonumber 
\label{elecP}
&& \frac{\partial p_{e}}{\partial t}+\left(\mathbf{u_{e}} \cdot \nabla\right)p_{e}= -\gamma_{e} p_{e}\left(\nabla\cdot\mathbf{u_{e}}\right) \\ \nonumber
&& \, +  \sum_{t=\mathrm{s,n}}\frac{\rho_{e}\nu_{et}}{m_{e}+m_{t}}\left[2k_B\left(T_{t}-T_{e}\right)+\frac{2}{3}m_{t}\left(\mathbf{u_{t}-u_{e}}\right)^{2}\right] \\ 
\nonumber 
&& \, - k_B\frac{\mathcal{L}_{e}T_{e}}{m_{e}} + \frac{2}{3}n_{n}(\nu_{ph,n}\mathcal{E}_{ns}^{exc}-\nu_{imp,n}\mathcal{E}_{ns}^{pot})  \\
&& \, - \frac{2}{3}n_e n_n\mathcal{R}_{en}^{inelastic} +\frac{1}{3}S_{e}\left(\mathbf{u_{n}-u_{e}}\right)^{2}\\ \label{Bind}
&&\frac{\mathbf{\partial B}}{\partial t}=\nabla\times(\mathbf{u_{+}}\times\mathbf{B}-\eta\mathbf{J}) 
\end{eqnarray}
where $\nu$ is the collision frequency between species, $\mathbf{u_+}$ is the charge-averaged velocity,
\begin{equation}
\mathbf{u_{+}}=\sum_{s=ions}\frac{q_{s}n_{s}u_{s}}{en_{e}}
\end{equation}
and $\eta$ is the magnetic diffusivity, which is defined as
\begin{equation}
\eta = \frac{1}{\mu_0\sigma_e} =  \frac{1}{\mu_0}\left(\frac{1}{\sigma_{en}}+\frac{1}{\sigma_{ei}} \right)
\end{equation}
where the electron conductivity, $\sigma_e$, comprises of contributions from electron-neutral ($\sigma_{en}=e^2n_e/\Sigma_{n'}\nu_{en'}m_e$) and electron-ion ($\sigma_{ei}=e^2n_e/\Sigma_{s'}\nu_{es'}m_e$) collisions.

In the above set of equations, note that the source ($\mathcal{S}$) and loss ($\mathcal{L}$) terms for species $s$ associated with photoionization ($\nu_{ph,s^{\prime}}$), electron impact ionization ($\nu_{imp,s^{\prime}}$), charge exchange ($k_{is^{\prime}}$), and recombination ($\alpha_{R,s}$) are described below:
\begin{eqnarray}\label{Ss}
\mathcal{S}_{s}&=&m_{s}n_{s^{\prime}}\left(\nu_{ph,s^{\prime}}+\nu_{imp,s^{\prime}}+\underset{i=\mathrm{ions}}{\sum}k_{is^{\prime}}n_{i}\right) \\
\mathcal{L}_{s}&=&m_{s}n_{s}\left(\alpha_{R,s}n_{e}+\underset{t^{\prime}=\mathrm{neutrals}}{\sum}k_{st^{\prime}}n_{t\prime}\right) \\
\mathcal{S}_{e}&=&m_{e}\underset{s'}{\sum}\,n_{s'}\,(\nu_{ph,s^{\prime}}+\nu_{imp,s^{\prime}}) \\
\mathcal{L}_{e}&=&m_{e}n_{e}\underset{s=\mathrm{ions}}{\sum}\alpha_{R,s}n_{s}
\end{eqnarray}

Inelastic collisions between electrons and neutral water molecules constitute an efficient way of cooling the electrons in the planetary lower ionosphere where collisions are highly frequent. Hence, we include the overall (rotational, vibrational, and electronic) cooling rate coefficient, $\mathcal{R}_{en}^{inelastic}$ (in eV cm$^3$s$^{-1}$),  in (\ref{elecP}) by following the prescription provided in \citet{gombosi15}:
\begin{eqnarray} \label{Qinel}
\mathcal{R}_{en}^{inelastic} &=& 4 \times 10^{-9} \left[1-\exp \left(-\frac{k(T_e-T_n)}{0.033 eV} \right)  \right] \\ \nonumber
&&+\, A \left[0.415-\exp \left(-\frac{kT_e-0.10eV}{0.10 eV} \right)  \right]
\end{eqnarray}
where $A=0$ for kT$_e$ $\le$ 0.188 eV and $A=6.5\times10^{-9}$ for kT$_e$ $>$ 0.188 eV.

In order to account for photoionization, we calculate the optical depth of the neutral atmosphere by applying the Chapman functions based on the numerical evaluation given by \citet{smith1972}. The photoelectron gains an excess energy $\mathcal{E}_{ns}^{exc}$ through the photoionization process \citep{HKL92} while it loses the ionization energy of H$_2$O during the electron impact ionization process \citep{Hay14} as indicated in Eq. (\ref{elecP}). The electron number density and velocity can be obtained by assuming quasineutrality \citep{Toth12}, 
\begin{eqnarray}
n_e = \frac{1}{e}\sum_{s=\mathrm{ions}}n_sq_s
\end{eqnarray}
and in terms of the current,
\begin{equation} \label{Ue}
\mathbf{u}_e = \mathbf{u_+} - \frac{{\mathbf{J}}}{e n_e} = \mathbf{u_+} - \frac{{\mathbf{\nabla\times B}}}{\mu_0 e n_e}
\end{equation}
where we have adopted Amp\'{e}re's law $\nabla\times \mathbf{B} = \mu_0 \mathbf{J}$, with $\mu_0$ denoting the vacuum permeability.

By using (\ref{ionmass}) - (\ref{Ue}), we are now in a position to simulate the water ions, the stellar wind protons, and the electrons in a fully self-consistent manner; both the individual behavior of each fluid and interactions between them have been taken into account. The array of processes included in this model are summarized below.
\begin{itemize}
    \item We have included the effects of both elastic and inelastic collisions. In Table \ref{rates}, the elastic collision rates between different fluid species are presented.
    \item The chemical reactions between species are manifested as source terms on the right hand side, including ionization (photoionization and electron impact ionization), charge exchange between neutrals and ions, and recombination. The electron impact ionization rates have been adopted from \citet{cravens87}. 
    \item Table \ref{reacoeff} summarizes the chemical reactions and the associated rates for inelastic collisions used in the multi-fluid MHD calculations.
\end{itemize}
A detailed discussion of the source terms and the inherent chemistry can also be found in \citet{rubin14b} and \citet{huang16a}.

\begin{table}
\caption{The elastic collision rates have been adopted from \citet{huang16a}. $Z_s$, $m_s$, $n_s$ and $T_s$ denote the charge state, mass (in amu), number density (in cm$^{-3}$) and temperature (in K) of a given species. Here, $m_{st}=\frac{m_sm_t}{m_s+m_t}$ and $T_{st}=\frac{m_sT_t+m_tT_s}{m_s+m_t}$ are the reduced mass and temperature. $C_{sn}$ represent numerical coefficients that have been presented in \citet{schunk2009}.}\label{rates}
\begin{center}
\begin{tabular}{c|c}
\hline
Elastic Collision Rates & Value (s$^{-1}$) \\
\hline
ion-ion ($\nu_{ii}$) & $1.27 \times \frac{Z_s^2 Z_t^2 \sqrt{m_{st}}}{m_s} \frac{n_t}{T_{st}^{3/2}}$ \\
ion-neutral ($\nu_{in}$) & $ C_{sn}n_n$ \\
electron-ion ($\nu_{ei}$) & $54.5 \times \frac{n_sZ_s^2}{T_e^{3/2}}$ \\
ion-electron ($\nu_{ie}$) & $1.27 \times \frac{\sqrt{m_e}}{m_s} \frac{n_sZ_s^2}{T_e^{3/2}}$ \\
electron-neutral ($\nu_{en}$) & $2.745 \times 10^{-5} n_n T_e^{-0.62}$ \\
\hline
\end{tabular}
\end{center}
\label{default}
\end{table}

\begin{table*}
\centering
\caption{Chemical reactions and associated rates for the ocean planets multi-fluid MHD code.}\label{reacoeff}
\begin{tabular}{llll}
\hline
\hline
\multicolumn{2}{c} {Chemical Reaction} & Rate (s$^{-1}$) \\
\hline
\multicolumn{3}{c} {Primary Photolysis and Particle Impact} \\
\hline
& H$_{2}$O + $h\nu$ $\rightarrow$  H$_2$O$^+$ + $e^{-}$   &  8.28 $\times$ 10$^{-7}$  s$^{-1}$\footnote{The photoionization rate represents the value at a distance of 1 AU from the Sun based on current solar cycle maximum conditions, i.e. the 1 EUV case in Table \ref{SW}, and must be appropriately rescaled when dealing with early epochs or other stellar systems.} & \citet{huebner2015} \\
 & e$^-$ + H$_{2}$O $\rightarrow$ e$^-$ + H$_2$O$^+$ + $e^{-}$   & see text & \citet{cravens87} \\
\hline
\multicolumn{2}{c} {Ion-Neutral Chemistry} & Rate (cm$^{3}$ s$^{-1}$) \\
\hline
&  H$_2$O$^+$ + H$_2$O $\rightarrow$  H$_2$O + H$_2$O$^+$  & $1.7 \times 10^{-9}$  &  \citet{Gom96} \\
&  H$^+$ + H$_2$O $\rightarrow$  H + H$_2$O$^+$  & $1.7 \times 10^{-9}$  &  \citet{Gom96} \\
\hline
\multicolumn{2}{c} {Electron Recombination Chemistry} & Rate (cm$^{3}$s$^{-1}$) \\
\hline
&										    & $1.57 \times10^{-5}T_e^{-0.569}$, $T_e \le 800 K$ & \\
&  H$_2$O$^+$  + $e^{-}$ $\rightarrow$  H$_2$O  & $4.73 \times10^{-5}T_e^{-0.74}$, $800 K < T_e \le 4000 K$ &  \citet{schunk2009} \\
&										    & $1.03 \times10^{-3}T_e^{-1.11}$, $T_e > 4000 K$ &  \\
& H$^+$  + $e^{-}$ $\rightarrow$  H  & $4.8 \times10^{-12}\left(\frac{250}{T_e}\right)^{0.7}$,  &  \citet{schunk2009} \\
\hline
\hline
\end{tabular}
\end{table*}

\section{Simulation set-up} \label{SecSim}

\begin{table*}
\caption{Three hypothetical stellar wind input parameters for a solar-type star based on: (i) typical solar wind parameters at 1 AU at the current epoch \citep{schunk2009}, (ii) ancient ($4.02$ Gyr) solar wind parameters at 1 AU \citep{Boss10} and (iii) solar wind values at the maximum total pressure of an extreme ``Carrington-type'' space weather event \citep{ngwira14}. Note that 1 EUV (below) refers to the EUV flux received at Earth during the solar cycle maximum.} \label{SW}
\centering
\begin{tabular}{llllll}
\hline
\hline
 & n$_{sw}$ (cm$^{-3}$) & v$_{sw}$ (km/s) & IMF (nT) & Radiation & H$_2$O$^+$ loss rate (s$^{-1}$) \\
\hline
Current & 8.7  & (-468, 0, 0)  & (-4.4, 4.4, 0) & 1 EUV & 6.7$\times$10$^{25}$ \\
\hline
Early & 136.7  &  (-910, 0, 0) & (-15.6, 30.2, 0) & 12 EUV & 6.0$\times$10$^{26}$  \\
\hline
Carrington Event & 424.5  &  (-1937.5, 6.7, -13.0) & (0, 23.0, -194.3) & 12 EUV & 7.3$\times$10$^{27}$  \\
\hline
\hline
\end{tabular}
\end{table*}

To this date, as noted in Sec. \ref{SecIntro}, no conclusive ocean planets have been detected, although several candidates have been identified. Consequently, there is a lack of concrete information regarding their atmospheric properties such as the composition and mass. Thus, for the purposes of this preliminary study, we consider the scenario where an Earth-sized ocean planet is orbiting a solar-type star. This situation roughly corresponds to Kepler-22b; Earth may also have been an ocean planet in the Hadean (or Archean) era, but its atmospheric composition is expected to have involved other gases.

In accordance with our choice, we employ the atmospheric temperature profile of Earth \citep{schunk2009}, and assume that its surface pressure is $1$ bar. We also specify a fiducial value for the magnetic dipole moment that is equal to that of Earth, which is a reasonable assumption for Earth-like exoplanets. As noted earlier, we also assume that water vapor is the primary neutral atmosphere component, and neglect other neutral species for the purposes of this study. The neutral atmospheric background adopted in our model incorporates the cold trap for water in the lower altitudes, but its effects on the ion losses are minimal due to the high-altitude source of the escaping ions. In future investigations, we will incorporate water vapor into primordial or current Earth/Mars atmospheres, and study the (ionized) water vapor escape rates for a wider range of exoplanets endowed with other neutral and ion species. 

The simulation domain starts at $100$ km, where H$_2$O$^+$ ion density satisfies the photochemical equilibrium condition and float boundary conditions for the velocity $\mathbf{u}$ and the magnetic field $\mathbf{B}$ are applied. The simulation box extends up to $200$ planetary radii with a highly non-uniform spherical grid. The lowest resolution is $10$ km in the planetary lower ionosphere, as this value is small enough to capture the scale height variation in the lower ionosphere and thermosphere. The horizontal (angular) resolution is 3$^{\circ}$ $\times$ 3$^{\circ}$. The rest of the details concerning the simulation setup are similar to those described in \citet{DLMC}.

The above set of equations in Sec. \ref{SSecMFMHD} is solved by using an upwind finite-volume scheme based on an approximate Riemann solver to ensure the appropriate conservation of plasma variables. Moreover, hyperbolic cleaning is adopted for divergence control. In order to determine the steady-state solution, the simulation starts with a two-stage local time-stepping scheme that enables different grid cells to select varying time steps based on the local stability condition, thereby accelerating the convergence to the steady-state and saving computational resources. Because of the stiffness of the source terms, a point implicit scheme is used for handling them \citep{Toth12}.

We will investigate three different cases in this paper. The first utilizes the solar wind conditions at the present Earth, while the other two cases simulate the atmospheric loss arising from the ancient solar wind around $4$ Gya, and an extreme  ``Carrington-type'' space weather event. The corresponding stellar wind parameters have been presented in Table \ref{SW}. We have chosen to work with the Planet-Star-Orbital (PSO) coordinates, where the X-axis points from the planet toward star, the Z-axis is perpendicular to the planetary orbital plane, and the Y-axis completes the right-hand system. In this study, we assume that the planetary dipole moment is along the Z-axis and has the same polarity as Earth.

\section{Results} \label{SecRes}

Our primary result concerns the magnitude of the ion escape rates for water ions, i.e. for $\mathrm{H_2O^+}$. We find that the escape rate for the current solar wind parameters is $6.7 \times 10^{25}$ s$^{-1}$, while the corresponding value at $4$ Gya is $6.0 \times 10^{26}$ s$^{-1}$ as shown in Table \ref{SW}. Hence, we find that the numerical escape rate $4$ Gya is a factor of $9.0$ higher than the current value. It is quite interesting that this factor appears to be in good agreement with the theoretical value of $8.7$ for unmagnetized or weakly magnetized planets; the theoretical prediction follows from $\dot{N}(t) \propto \dot{M}_\star \propto t_s^{-\alpha}$ with $\alpha \sim 1$ \citep{John15,LiLo17}.  

It is well-known that space weather events, such as coronal mass ejections (CMEs), occur frequently on young Sun-like stars \citep{Shi13}. Hence, we have also investigated an extreme ``Carrington-type'' space weather event \citep{ngwira14} to approximately mimic a typical CME from a stormy young Sun. The ion escape rate reaches $7.3 \times 10^{27}$ s$^{-1}$, about one order of magnitude higher than the normal ancient case ($6.0 \times 10^{26}$ s$^{-1}$). Our analysis suggests that such space weather events may prove to be a key driver of atmospheric losses for (exo)planets orbiting an active young Sun-like star.

\begin{figure*}[!ht]
\centering
\includegraphics[scale=1.0]{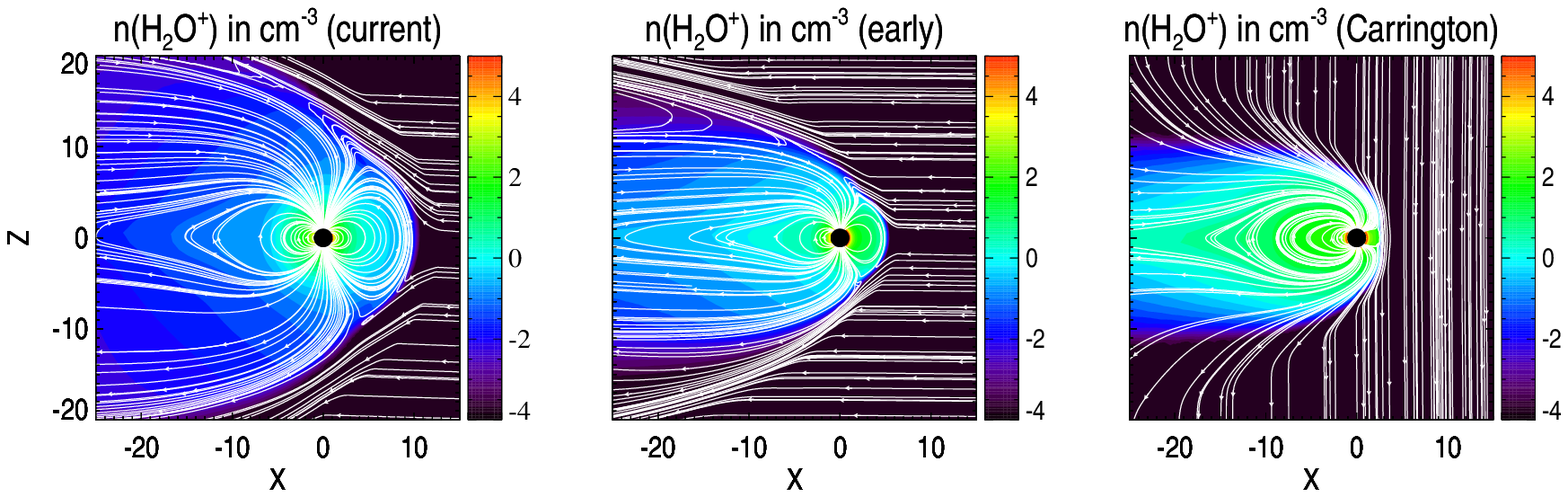}
\includegraphics[scale=1.0]{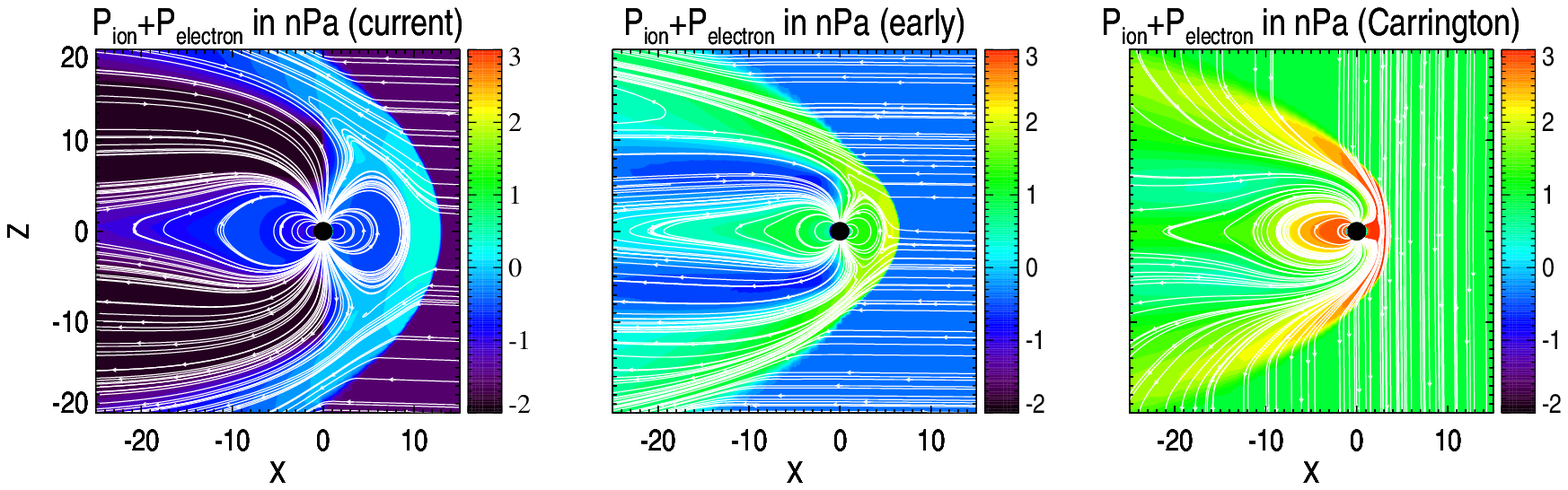}
\caption{The logarithmic scale contour plots of the H$_2$O$^+$ ion density (first row) and the sum of the ion (P$_{ion}$) and electron (P$_{electron}$) pressure (second row) with magnetic field lines (in white) in the meridional plane for current solar wind conditions, early solar wind conditions and an extreme ``Carrington-type'' space weather event.}\label{contour}
\end{figure*}

\begin{figure}
\centering
\includegraphics[scale=0.42]{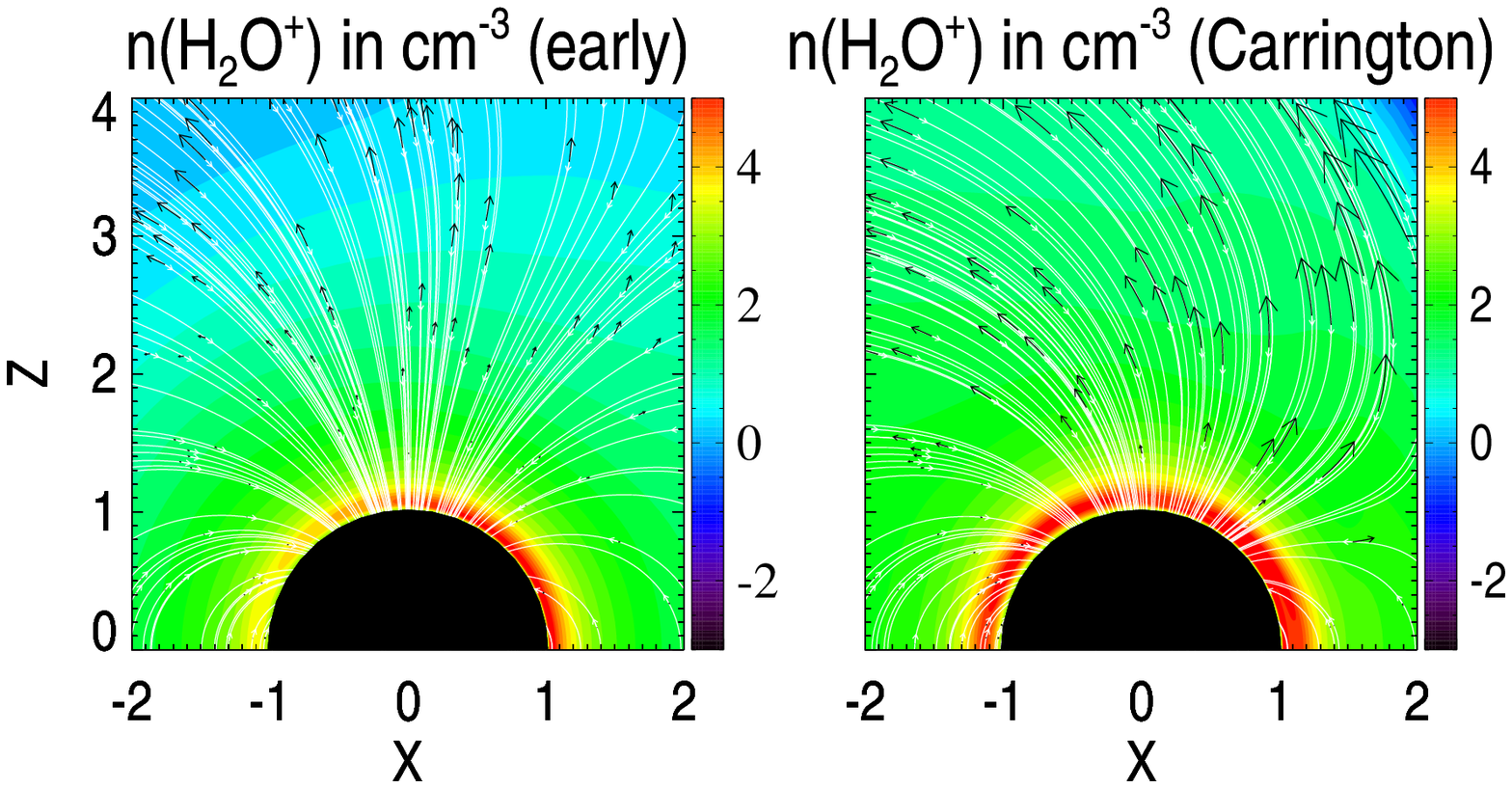}
\includegraphics[scale=0.42]{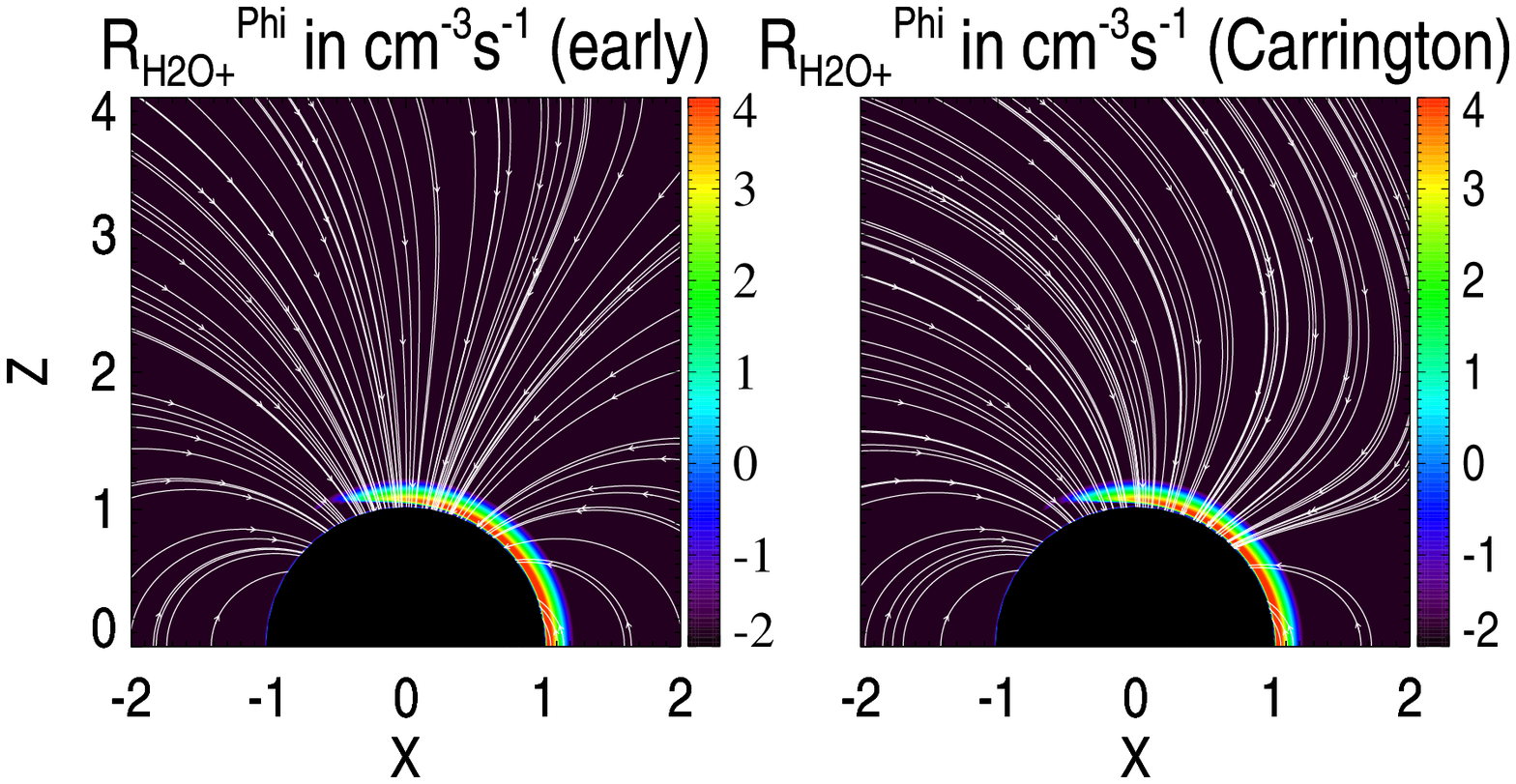}
\includegraphics[scale=0.42]{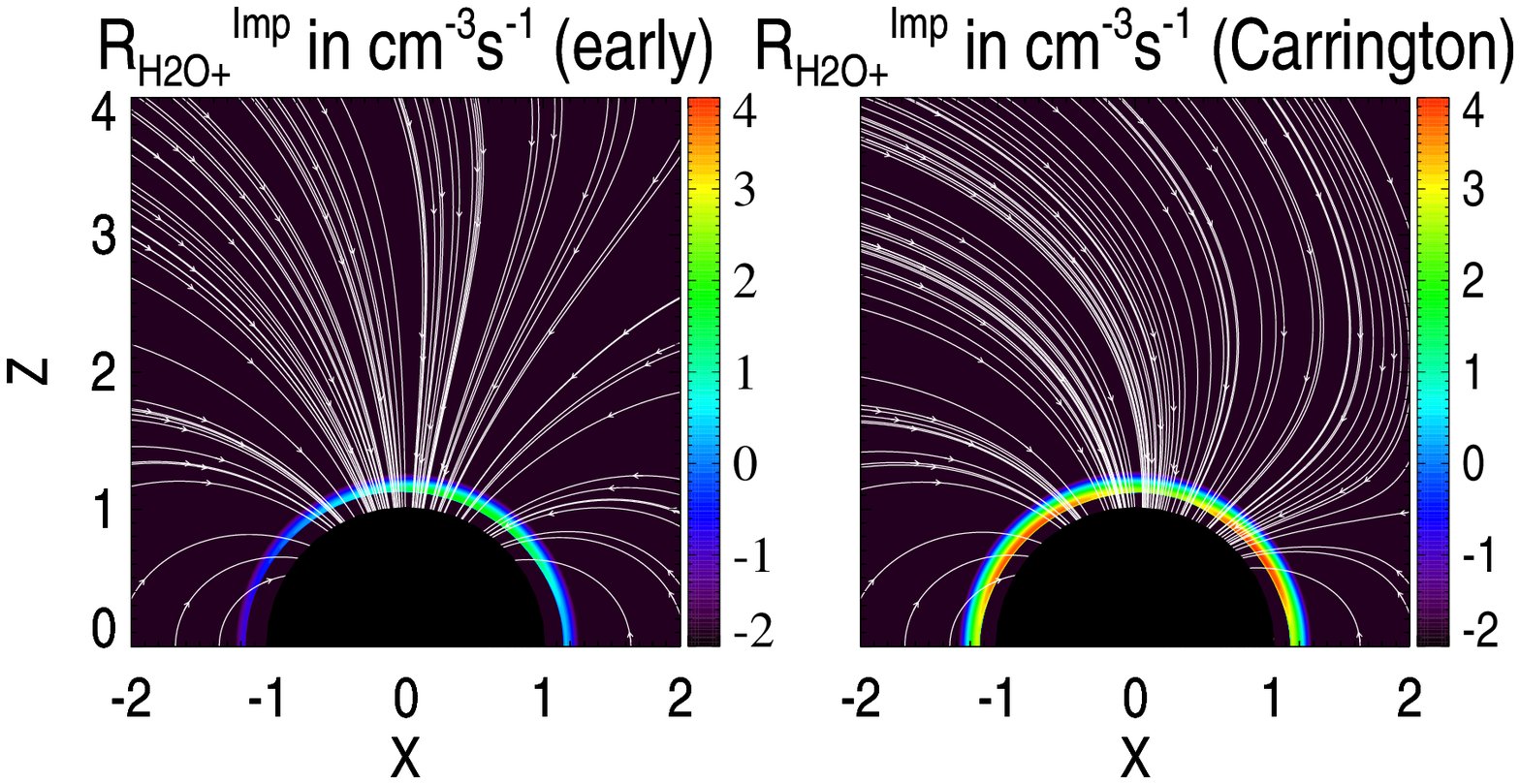}
\includegraphics[scale=0.42]{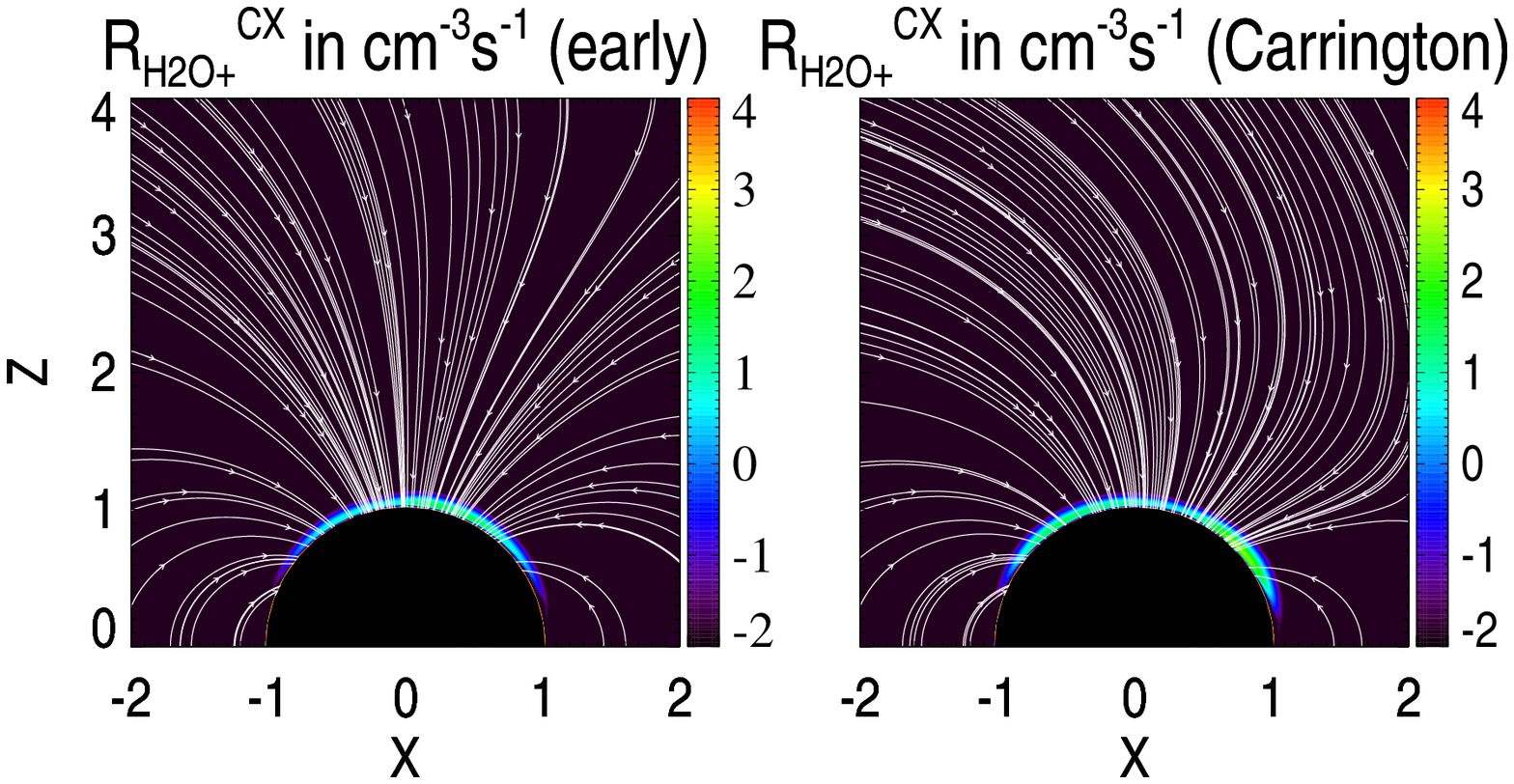}
\caption{First row: The logarithmic scale contour plots of the H$_2$O$^+$ ion density with H$_2$O$^+$ ion velocity vectors (in black) and magnetic field lines (in white) in the meridional plane for early solar wind conditions and an extreme ``Carrington-type'' space weather event. Note that the black arrows depict both the direction and the magnitude of H$_2$O$^+$ ion velocities. The acceleration of H$_2$O$^+$ ions is demonstrated by the variations in the vector magnitude with altitudes. Second to fourth rows: The logarithmic scale contour plots of the photoionization rate R$_{H2O+}^\mathrm{Phi}$, electron impact ionization rate R$_{H2O+}^\mathrm{Imp}$, and charge exchange rate R$_{H2O+}^\mathrm{CX}$ (with stellar wind protons) of H$_2$O$^+$ ions.}\label{contour2}
\end{figure}

Fig. \ref{contour} presents the contour plots of the H$_2$O$^+$ ion density, the sum of the ion and electron thermal pressures and the magnetic field lines for the three cases. The planetary magnetosphere is significantly compressed from left to right due to the enhanced stellar wind dynamic pressure (P$_{dyn}=n_{sw}m_{sw}v_{sw}^2$). An inspection of Fig. \ref{contour} reveals that H$_2$O$^+$ ions escape from the planet along open geomagnetic field lines through the lobes and magnetotail. The outflow of H$_2$O$^+$ ions (polar wind) is mainly caused by their acceleration (Fig. \ref{contour2}, first row) resulting from the electron pressure gradient, $\nabla p_e$ (i.e., the ambipolar electric field), the $\mathbf{J}$ $\times$ $\mathbf{B}$ force, and further pickup by the stellar wind at high altitudes due to the Lorentz force, $n_{s}q_{s}\left(\mathbf{u_{s}-u_{+}}\right)\times\mathbf{B}$, in the ion momentum equation (\ref{ionmom}); the latter exists only in the multi-fluid MHD model since it vanishes for the multi-species MHD model \citep{dong14}. As seen from the second row of Fig. \ref{contour}, a stronger stellar wind dynamic pressure is commensurate with larger plasma thermal pressure in the magnetosheath, and is consistent with the magnetospheric shape.

In Fig. \ref{contour2}, the smooth transition of the photoionization rate around the terminator is a result of adopting the Chapman function \citep{smith1972}. It is noteworthy that the nightside H$_2$O$^+$ ions are mainly produced by electron impact ionization, as seen from equation (\ref{Ss}) and Table \ref{reacoeff}. The contribution of charge exchange with stellar wind protons is concentrated around the polar cap regions due to the proton precipitation along the open magnetic field lines; the charge exchange with H$_2$O$^+$ has not been plotted explicitly.

\section{Conclusions}
We have developed a numerical code capable of accurately simulating the water ions escaping from (exo)planets. Our code contains the relevant atmospheric chemistry, and evolves each species independently in accordance with the multi-fluid MHD equations. We applied this model to a water world with planetary and stellar wind parameters equal to that of the current and ancient ($4$ Gya) Earth-Sun system. 

We found that the ion escape rate of $\mathrm{H_2O^+}$ from this planet at the ``current'' epoch is $6.7 \times 10^{25}$ s$^{-1}$. This value is slightly higher than, or comparable to, the ion escape rates observed for the terrestrial planets in our Solar system \citep{Lammer13}. We also demonstrated that the escape rate at $4$ Gya was about an order of magnitude higher than the present-day value. For an extreme ``Carrington-type'' space weather event that is expected to have been frequent in the Hadean epoch, the $\mathrm{H_2O^+}$ ion escape rate can reach $\sim 10^{28}$ s$^{-1}$, about two orders of magnitude higher than the current Earth value. In other words, for active stars (of any spectral type), space weather events must be taken into account when studying the effects of ion atmospheric losses.

At this stage, we wish to reiterate a few caveats. The planetary and stellar parameters chosen herein were based on fiducial values, and must be duly altered when studying a given exoplanet-star interaction. Second, the atmosphere was assumed to be consist mostly of water vapor, although we intend to generalize our model to include other species in the future. Lastly, any statements concerning atmospheric depletion must be interpreted with caution since sources like outgassing and impacts were not taken into account. 

For all cases considered herein, we find that Earth-like oceans (with a total mass of $\sim 10^{24}$ g) will not be evaporated over Gyr timescales as the ion escape rates are far too low (by $3$-$4$ orders of magnitude). In contrast, for exoplanets in the HZ of M-dwarfs, the situation is very different because of the fact that the escape rate approximately scales with the inverse of the distance \citep{LiLo17}. Hence, the escape rate (of water ions) will be $2$-$3$ orders of magnitude higher than on Earth even without considering space weather effects. Thus, planets orbiting these stars could have their oceans depleted over Gyr timescales, especially when extreme space weather events are taken into account. 

Rapid desiccation would have important consequences for the evolution of life on such planets, especially with regards to oceanic and coastal biodiversity, productivity, and food webs. Another important consequence of desiccation is that water has been identified as one of the crucial ingredients for the functioning of plate tectonics. If plate tectonics were to shut down, a wide array of deleterious effects would follow \citep{Lam09}. One of the most notable among them is that the planetary dynamo could shut down, and unmagnetized planets are more vulnerable to sources of ionizing radiation and atmospheric stripping by the stellar wind. 

Hence, we conclude by observing that future studies of planetary habitability should take the effects of water ion losses into account.

\acknowledgments
The authors acknowledge fruitful discussions with Adam Burrows and Abraham Loeb. CD is supported by the NASA Living With a Star Jack Eddy Postdoctoral Fellowship Program, administered by the University Corporation for Atmospheric Research. ML was supported by the Institute for Theory and Computation (ITC) at Harvard University. Resources for this work were provided by the NASA High-End Computing (HEC) Program through the NASA Advanced Supercomputing (NAS) Division at Ames Research Center. The Space Weather Modeling Framework that contains the BATS-R-US code used in this study is publicly available from \url{http://csem.engin.umich.edu/tools/swmf}. For distribution of the model results used in this study, please contact the corresponding author.


\end{document}